\documentclass[aps,prb,twocolumn,groupedaddress]{revtex4}
\usepackage{graphicx}
\usepackage{wasysym}
\usepackage{bm}

\begin{document}
\title{ Effect of the edge states on the conductance and thermopower
in Zigzag Phosphorene Nanoribbons}
\author{R. Ma$^{1,2\footnote{These authors contributed equally to this work.}}$}
\email{njrma@hotmail.com}
\author{H. Geng$^{2,*}$}
\author{W. Y. Deng$^2$}
\author{M. N. Chen$^2$}
\author{L. Sheng$^{2,3}$}
\email{shengli@nju.edu.cn}
\author{D. Y. Xing$^{2,3}$}
\email{dyxing@nju.edu.cn}
\address{$^1$ School of Physics and Optoelectronic Engineering,
Nanjing University of Information Science and Technology, Nanjing
210044, China\\
$^2$ National Laboratory of Solid State Microstructures and
Department of Physics, Nanjing University, Nanjing 210093, China\\
$^3$ Collaborative Innovation Center of Advanced Microstructures,
Nanjing 210093, China}

\begin{abstract}
We numerically study the effect of the edge states on the conductance
and thermopower in zigzag phosphorene nanoribbons (ZPNRs) based on
the tight-binding model and the scattering-matrix method.
It is interesting to find that the band dispersion, conductance,
and thermopower can be modulated by applying a bias voltage and
boundary potentials to the two layers of the ZPNRs.
Under the certain bias voltage, the two-fold degenerate
quasi-flat edge bands split perfectly. The conductance can be
switched off, and the thermopower around zero energy increases.
In addition, when only the boundary potential of the top layer
or bottom layer is adjusted, only one edge
band bends and merges into the bulk band. The first conductance
plateau is strongly decreased to $e^2/h$ around zero energy.
Particularly, when the two boundary potentials are adjusted,
all the edge bands bend and
fully merge into the bulk band, and the bulk energy gap
is maximized. More interestingly, a pronounced conductance plateau
with $G=0$ is found around zero energy, which is attributable
to the opening of the bulk energy gap between the valence
and conduction bands. Meanwhile, the thermopower can be enhanced
more than twice, compared to that of the perfect ZPNRs.
The large magnitude of thermopower is ascribed to
the appearance of the bulk energy gap around zero energy.
Our results show that the modulated ZPNRs are more reliable
in thermoelectric application.
\end{abstract}

\maketitle

\section{Introduction}
\label{sec:intro}

Two-dimensional (2D) layered crystal materials have attracted
considerable attention due to their unique electric properties~\cite{Liao2010,Schwierz2010,Wu2011,Mak2010,Radisavljevic2011,Yoon2011}.
Graphene is known to possess excellent electronic properties such as high
carrier mobility and tunable carrier type and density, but its zero band gap limits its performance
~\cite{Liao2010,Schwierz2010,Wu2011}.
Transition-metal dichalcogenide MoS$_2$ is found to be a direct
band gap semiconductor, but the carrier mobility can only reach
200 $cm^2/V$$\cdot s$~\cite{Mak2010,Radisavljevic2011,Yoon2011}.
Very recently, another new semiconducting material,
called black phosphorus, has attracted much attention because of
its great transport properties and potential applications
~\cite{Reich2014,Li2015,Liu2014,Qiao2014,Xia2014,Koenig2014,Gomez2014,Li2014}.
Black phosphorus is a layered material, in which individual
atomic layers are stacked together by van der Waals interactions,
and the phosphorene is another 2D stable elemental material
that can be mechanically exfoliated from bulk black phosphorus~\cite{Li2014,Liu2014}.
Within a phosphorene sheet, every phosphorous atom is covalently bonded
with three neighboring atoms forming a highly corrugated honeycomb-like
structure with low symmetry and high anisotropy. Unlike the zero-gap
semimetal graphene, phosphorene exhibits a direct band gap that can be
modified from 1.51 $eV$ of a monolayer to 0.59 $eV$ of
five layers~\cite{Qiao2014}.
Experiments~\cite{Li2015} have shown that black phosphorus exhibits
an integer Quantum Hall Effect.
Field effect transistors based on few layers of phosphorene have been
reported to show high performance at room temperatures with an on/off ratio
up to $10^5$ and a much larger mobility than MoS$_2$, up to 1000 $cm^2/V\cdot s$~\cite{Li2014,Liu2014,Gomez2014,Xia2014},
making it a promising candidate material for electronic device applications.
On the other hand, experimental measurements~\cite{Flores2015} of the
thermoelectric power of bulk black phosphorus demonstrated that
its Seebeck coefficient at room temperatures is 335$\pm 10\mu V/K$,
and the thermopower increases with temperature,
suggesting that black phosphorus also has great potential in
thermoelectric applications~\cite{Qin2014,Lv2014,Fei2014,Peng2013}.

One-dimensional (1D) nanoribbons etched or patterned from above
2D materials can offer even more tunability in electronic
properties because of the unique quantum confinement and edge effects.
The electronic properties of the phosphorene nanoribbons (PNRs), especially the zigzag
phosphorene nanoribbons (ZPNRs), have been studied~\cite{Tran2014,Ezawa2014,Sisakht2015,Guo2014,Han2014}.
It was found that the pristine ZPNRs are metals regardless
of the ribbon widths, while the pristine Armchair PNRs (APNRs)
are semiconductors with indirect band gaps~\cite{Guo2014}.
The band gap and the transport
properties of the PNRs can be dramatically modified by tensile strain
and external in-plane electric-field~\cite{Guo2014,Han2014,Ezawa2014,Sisakht2015}.
The possible structural reconstruction in the edge of the PNRs has
also been investigated~\cite{Carvalho2014}.
More recently, theoretical studies find intriguing properties in
skewed-ZPNRs and skewed-APNRs. The skewed-APNRs is particularly
interesting since it has two quasi-flat bands in the middle of
the band gap for the band structure. They share the same topological
origin as those in the normal ZPNRs~\cite{Peeter2016}.
On the other hand, the first-principle calculations~\cite{Zhang2014}
showed that in the ZPNRs, the Seebeck coefficient around the Fermi level
exhibits very small value, but can be greatly enhanced by hydrogen
passivation at the edge. The resulting giant Seebeck coefficient
is very beneficial for their thermoelectric applications.
In the experiment, Chang-Ran Wang $et$ $al.$~\cite{Wang2011} demonstrate
that the thermopower can be enhanced greatly at a low temperature by using
a dual-gated bilayer graphene device, which has been predicted
theoretically and originates from the opening of a band gap~\cite{hao2010}.
Up to now, there are no experimental studies on tuning the thermopower
of the ZPNRs.
Owing to the unique structure and edge effect, the ZPNRs are
expected to exhibit rich novel electric and thermoelectric transport
properties under external electric field.
However, theoretical studies of the electric and thermoelectric
transport of the ZPNRs are limited, compared with
those of other 2D materials~\cite{Zhu2010,Ma2011,Fan2014,Peng2016}.
In particular, a detailed investigation about the effect of
the edge states on the conductance and thermopower of the ZPNRs,
in the presence of an applied bias voltage,
has not been carried out so far, which is highly desired.
Such theoretical study will provide useful theoretical predicting and
guidance to the experimental research of the electric and thermoelectric
transport in these systems.

In this paper, we perform numerical study of the effect of
edge states on the conductance and thermopower in the ZPNRs based on the
tight-binding model and scattering-matrix method.
We find that the edge band dispersion, conductance, and thermopower
can be modulated by applying an electrostatic bias voltage
between the top and bottom layers and boundary potentials
to the layers of the ZPNRs.
Under certain bias voltage, the two-fold degenerate
quasi-flat edge band splits completely. The first conductance
plateau around zero energy becomes $0$, compared to that
of the perfect ZPNRs. The thermopower around zero energy can be
enhanced by increasing the bias voltage.
When only the boundary potential of the top layer or bottom
layer is adjusted, only one edge
state bends and merges into the bulk band.
Around zero energy, the first conductance plateau is strongly
reduced to $e^2/h$. Particularly, when the two boundary
potentials are adjusted, all the edges states bend and
fully merge into the bulk band
and the bulk energy gap is maximized. More interestingly,
a pronounced plateau with $G=0$ emerges around zero energy,
which can only be understood as due to the opening of the
bulk energy gap between the valence and conduction bands.
Meanwhile, the thermopower is enhanced more than twice,
compared to that of the perfect ZPNRs. We attribute
the large magnitude of thermopower to the appearance of
the bulk energy gap around zero energy.

This paper is organized as follows. In Sec.\ II, the model
Hamiltonian for zigzag phosphorene nanoribbons is introduced.
In Sec.\ III, numerical results from the exact calculations
based on the Landauer-B\"uttiker are presented. The final section
contains a summary.

\begin{figure}[tbh]
\par
\includegraphics[width=3.3in]{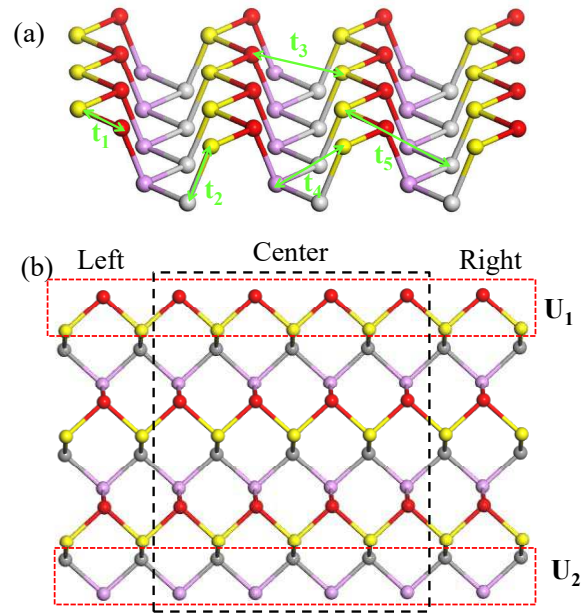}
\caption{(color online). (a) Crystal structure and hopping integrals
$t_i$ in the tight-binding model of phosphorene.
(b) The zigzag phosphorene nanoribbon with two terminals
from top view, and the boundary potentials of the top and bottom layers
are denoted by $U_{1}$ and $U_{2}$, respectively. Here, the red (yellow)
balls represent phosphorus atoms in the top layer, and
the gray (purple) balls represent phosphorus atoms in the bottom layers, respectively.
} \label{fig.1}
\end{figure}

\begin{figure*}[tbh]
\par
\includegraphics[width=0.8\textwidth]{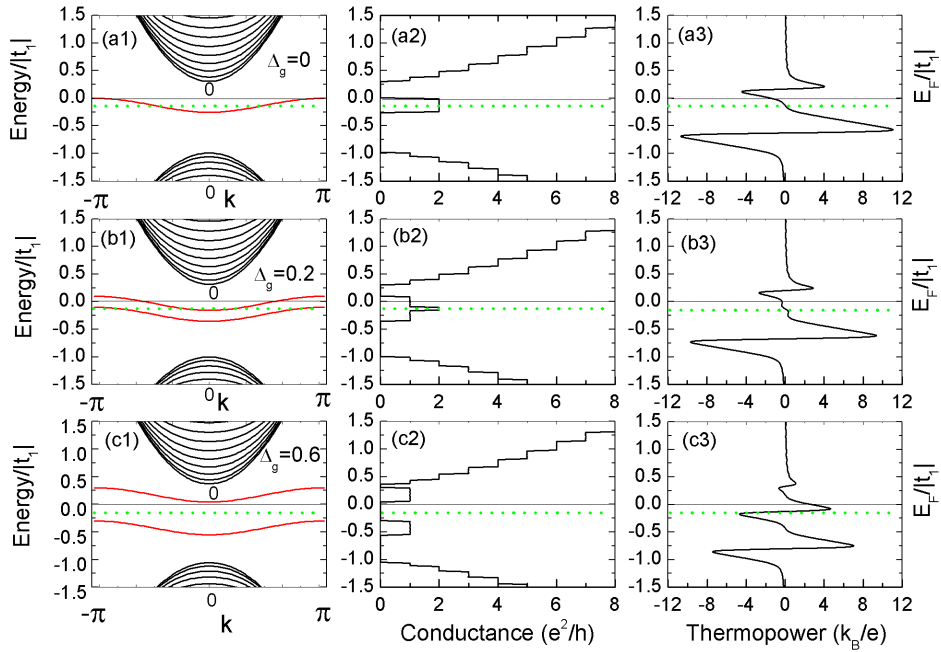}
\caption{(color online). Calculated band structure, conductance and thermopower
of the ZPNRs in the presence of an applied bias voltage $\Delta_g$.
(a) $\Delta_g=0$, (b) $\Delta_g=0.2\vert t_1 \vert$, and
(c) $\Delta_g=0.6\vert t_1 \vert$.
Red curves represent the quasi-flat bands, and green lines are the Fermi energy.
The width of the ZPNRs is set to $N=20$.
The temperature is taken to be $k_BT=0.03\vert t_1 \vert$ in the thermopower calculation.
} \label{fig.2}
\end{figure*}

\section{Model and Methods}
\label{sec:model}

The unit cell of phosphorene contains four phosphorus atoms,
where two phosphorus atoms sit on the top layer and the other
two sit on the bottom layer. The tight-binding model
Hamiltonian of phosphorene~\cite{Rudenko2014} with a bias voltage
can be described by
\begin{eqnarray}
H&=&\sum\limits_{\langle {ij}\rangle}t_{ij}c_{i}^{\dagger
}c_{j}+\sum\limits_{i}
U_i c_{i}^{\dagger}c_{i}+H.c.\ .
\end{eqnarray}
Here, the summation of $\langle {ij}\rangle$
runs over all neighboring lattice sites with
$t_{ij}$ as the hopping
integrals, and $c_{i}^{\dagger}$ and $c_{i}$
are the creation and annihilation operators of electrons
on site $i$. The hopping integrals between
a site and its neighbours are illustrated in Fig.\ref{fig.1}(a).
The hopping integral $t_{1}$ corresponds to the connection
in each zigzag chain in the upper or lower layer,
and $t_{2}$ stands for the connection between a pair of
zigzag chains in the upper and lower layers. $t_3$ is between the nearest-neighbour
sites of a pair of zigzag chains in the upper or lower layer,
and $t_4$ is between the next nearest-neighbour sites of
a pair of zigzag chains in the upper and lower layers. $t_5$ is the hopping between
two atoms on upper and lower zigzag chains that are farthest from each other.
The values of these hopping integrals are
$t_1 = -1.220$ $eV$, $t_2 =3.665$ $eV$, $t_3 =-0.205$ $eV$, $t_4 = -0.105$ $eV$,
and $t_5 = -0.055$ $eV$~\cite{Rudenko2014}.
The system has
negative hopping integral $t_1$ along the zigzag chains, and positive
hopping integral $t_2$ connecting the zigzag chains.
In the presence of a uniform perpendicular electric field,
the top and bottom layers gain different
electrostatic potentials $U_{top}$ and $U_{bottom}$, which are
taken to be antisymmetric, namely,
$U_{top}=-U_{bottom}=\frac{1}{2}\Delta_g$~\cite{Ma2011,Ma2009}.
If we choose a relatively small potential $\Delta_g$=0.1$\vert t_1 \vert$,
the realistic magnitude of the electric field is about 0.29 $V/nm$.
This is experimentally feasible~\cite{zhang2009}.

We consider a zigzag PNR with two terminals, as shown in Fig.\ref{fig.1}(b).
While a system with $N=3$ is shown in the schematic view,
much larger sizes are adopted in the following actual numerical calculations.
The linear conductance is calculated by using
the Landauer-B\"uttiker formula~\cite{Landauer1970,Datta1995,Lv2012}
\begin{eqnarray}
G &=& {2e^2\over h} \int dE
\left (-{\partial f_0
\over \partial E }\right) T_{LR}(E)\ ,
\label{eq:conductance}
\end{eqnarray}
where $f_0 = 1/[e^{(E-E_F)/k_B T}+1]$ is the
Fermi distribution function. $T_{LR}(E)$ is the transmission coefficient,
which can be calculated by using the Kwant package~\cite{Groth2014}, with
the superscripts $L$ and $R$ indicating the left and right leads,
respectively. The thermopower $S$ can be calculated by
using the formula~\cite{Lv2012}
\begin{eqnarray}
S&=&-\frac {k_B}{e} \frac
{\int dE
\left (-{\partial f_0
\over \partial E }\right) \frac{E-E_F}{k_B T} T_{LR}(E)}{\int dE
\left (-{\partial f_0
\over \partial E }\right) T_{LR}(E)}.
\label{eq:thermoelectric}
\end{eqnarray}

\begin{figure}[tbh]
\includegraphics[width=3.0in]{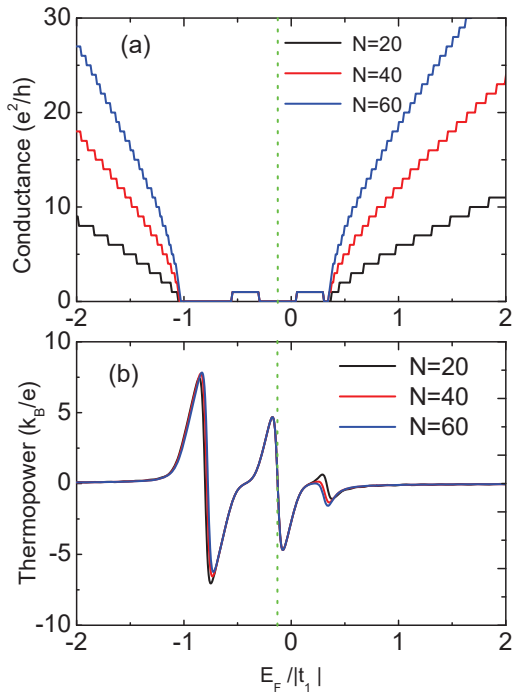}
\caption{ (color online). Calculated conductance and thermopower of the ZPNRs
under bias voltage $\Delta_g=0.6\vert t_1 \vert$ for three different ribbon widths.
The temperature is taken to be $k_BT=0.03\vert t_1 \vert$ in the thermopower calculation.
} \label{fig.3}
\end{figure}

\section{Results and Discussion}

The band structure and conductance of the ZPNRs at zero temperature
obtained by numerical diagonalization of the tight-binding Hamiltonian
are plotted in Fig.\ref{fig.2}. The results of
the perfect ZPNRs without applied bias voltage
are shown in Figs.\ref{fig.2}(a).
As can be seen from Fig.\ref{fig.2}(a1), the upper and lower
quasi-flat edge bands around the Fermi level are detached from
the bulk bands, and are two-fold degenerate. The wave functions of
the quasi-flat bands are always localized near the sample edges.
The conductance shown in Fig.\ref{fig.2}(a2) displays
a clear stepwise structure. When the Fermi energy
crosses the discrete transverse channels, the quantized
transmission coefficient jumps from one step to another.
The quantized conductance plateaus follow the sequence $ G =n e^2/h $
with $n$ as positive integers. The first conductance
plateau at the Fermi energy is $2 e^2/h $.
The conductance plateaus in the conduction band or valence
band are not equidistant in energy. The widths of conductance
plateaus are determined by the energy scale between the successive
modes in the energy spectrum, which depends on the ribbon width
and energy interval. Our calculated results are in good agreement
with those obtained by Ezawa $et$ $al.$~\cite{Ezawa2014,Sisakht2015}.
However, when a bias voltage or a potential difference $\Delta_g$ is
applied to the top and bottom phosphorene sheets, the band structure
and conductance show remarkable different features.
In Figs.\ {\ref{fig.2}(b)-(c), we show the
calculated band structure and conductance for the  ZPNRs
in the presence of an applied bias voltage.
As seen from Figs.\ref{fig.2}(b1) and (b2),
the potential difference between the top and bottom phosphorene sheets
lifts the degeneracy of the two edge modes, shifting one edge mode
upward and the other downward without changing their shapes.
The original first conductance plateau  $2 e^2/h $ still exists,
but its width decreases,
and a new conductance plateau of $e^2/h $ appears around the Fermi energy.
When the bias voltage increases to $\Delta_g=0.6\vert t_1 \vert$,
the upper and lower quasi-flat edge bands split completely, and the
minimal conductance becomes $0$ around the Fermi energy,
as shown in Figs.\ref{fig.2}(c1) and (c2).
This feature suggests that
the biased ZPNRs can serve as field-effect
transistors, whose conductance can be switched on and off by
a potential difference between the top and bottom phosphorene sheets.

We further calculate the thermopower $S$ using Eq.\ (\ref{eq:thermoelectric}),
which can be directly determined in experiments by measuring
the responsive electric fields. In the right side of the Fig.\ref{fig.2},
we show the calculated thermopower $S$ for some different bias voltages.
According to the definition of the thermopower, $S$ is determined
by the electron-transmission-weighted average value of the heat energy $E$-$E_F$.
In Fig.\ref{fig.2}(a3), for metallic ZPNRs without the bias voltage,
only the carriers within several $k_BT$ around the Fermi energy
are important for transport. Thus the average heat energy is
in the order of $k_BT$, which results in a much smaller thermopower.
However, when a bias voltage or a potential difference $\Delta_g$
is applied to the top and bottom phosphorene sheets, the peak value
of $S$ around zero energy increases.
With the increase of $\Delta_g$, the peak height of $S$
increases to $\pm 4.69 k_B/e$ ($\pm 404.14 \mu V/K$)
at $\Delta_g=0.6\vert t_1 \vert$, as shown in Fig.\ref{fig.2}(c3).
Around zero energy point, the sign of thermopower changes abruptly.
This is because near zero energy point the transmission coefficient
$T_{LR}$ is zero and the carriers cannot be transmitted.
The large magnitude of the thermopower is mainly due to the split
of the two-fold degenerate quasi-flat edge bands and the existence
of the energy gap in biased ZPNRs, so that the electrons
at the conduction subband edge, which are responsible for electron
conduction at the maximum thermopower, have a much larger $E$-$E_F$
compared to the case of without applied bias voltage.
This is similar to the corresponding behavior in semiconducting
armchair graphene nanoribbons~\cite{ouyang2009,xing2009,hao2010}.
The maximum magnetitude of thermopower, which is reached with
Fermi energy near the middle of the band gap.
The greatly enhanced thermopower is very beneficial for the
thermoelectric applications of the ZPNRs.

Next, we study the effect of the ribbon width on
the conductance and thermopower of the ZPNRs.
In Fig.\ref{fig.3}, we show the conductance at zero temperature and
thermopower for three different ribbon widths at fixed bias voltage
$\Delta_g=0.6\vert t_1 \vert$.
As we see from Fig.\ref{fig.3}(a),
with the increase of ribbon width, the characteristics of
the conductance are qualitatively similar. Around zero energy, the conductance
plateaus remain unchanged, which are independent of
the width of the ribbon.
Moreover, the widths of the other conductance plateaus away from
zero energy become smaller for $E>0.3\vert t_1 \vert$ and
$E< -1.0\vert t_1 \vert$ with increasing the ribbon widths.
As seen from Fig.\ref{fig.3}(b), the peak value of $S$ around
zero energy is nearly unchanged with the increase of the
ribbon width. Therefore, we can conclude that the characteristic
features of the conductance and thermopower around zero energy
are insensitive to the ribbon width.

\begin{figure}[tbh]
\includegraphics[width=3.5in]{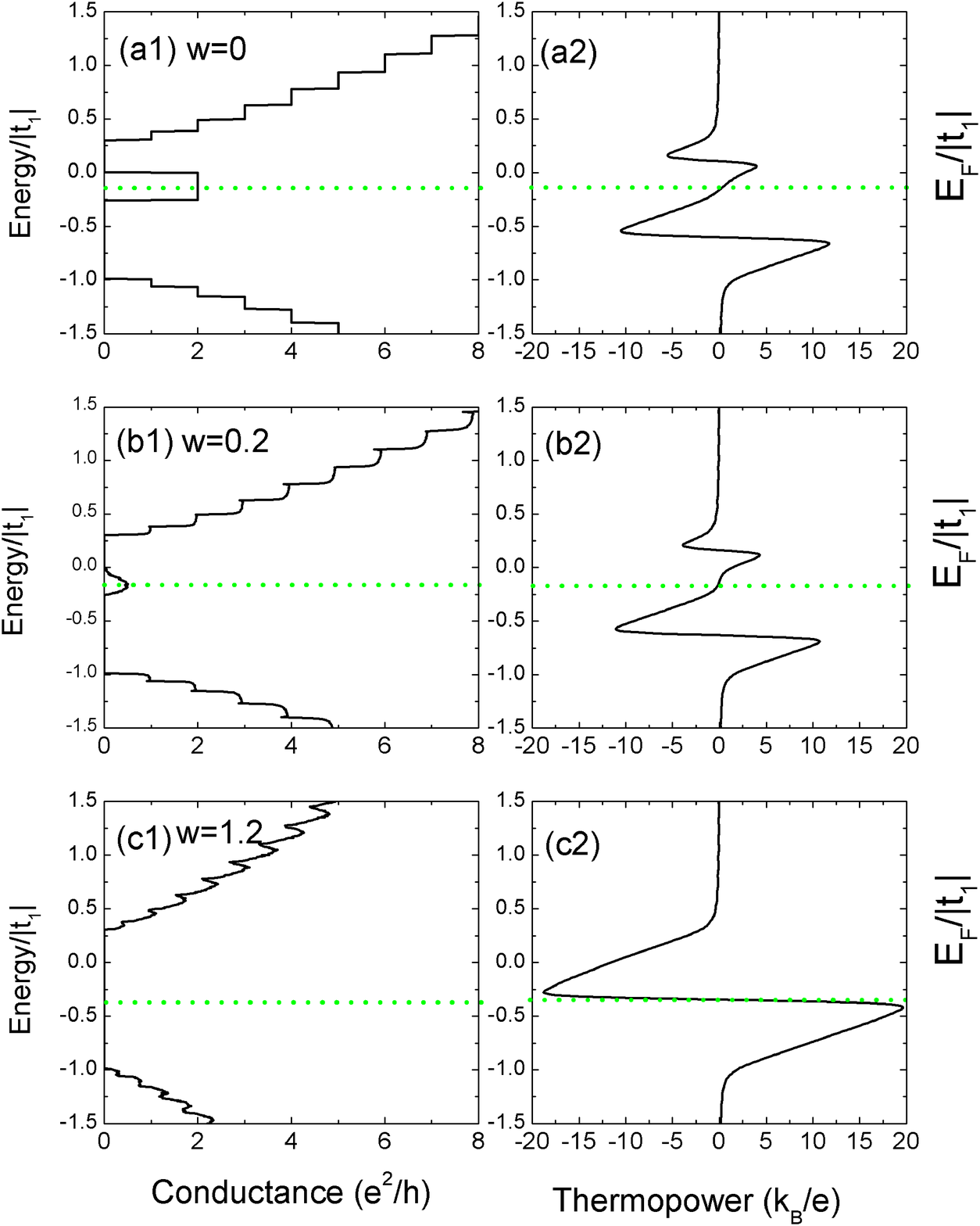}
\caption{ (color online). Calculated thermopower of the ZPNRs
without bias voltage for three different disorder strength. (a)W=0, (b)W=0.2 and (c)W=1.2.
The temperature is taken to be $k_BT=0.03\vert t_1 \vert$ in the thermopower calculation.
Each data point is obtained
by averaging over up to $2000$ disorder configurations.
} \label{fig.4}
\end{figure}

We also study the disorder effect on the conductance and thermopower
in ZPNRs without applied bias voltage. In our numerical calculation,
we consider the on-site Anderson disorder, and the on-site potential
energy $w_i$ is assumed to distribute uniformly in the region of
$w_i\in \lbrack -W/2,W/2]\vert t_1 \vert$, with disorder strength $W$.
Here, the random on-site potential is considered only in the central
scattering region. In Fig.\ref{fig.4}, the conductance and thermopower
are shown as a function of $E_F$ for three different disorder strengths.
For comparison, the calculated conductance $G$ and thermopower $S$
for a clean ZPNRs are plotted in Fig.\ref{fig.4}(a).
As the increase of the disorder, it is found that the conductance and
thermopower are strongly affected. As can be seen from Fig.\ref{fig.4}(b),
the original first conductance plateaus $G=2 e^2/h$ at the Fermi energy
begins to collapse even in the small disorder.
In Fig.\ref{fig.4}(c), at a relatively strong-disorder strength $W=1.2$,
the first conductance plateaus with $G=2 e^2/h$ disappears completely,
and a new conductance plateau with $G=0$ is found,
which can only be understand as due to the opening of energy gap
between the valence and conduction bands.
So, the peak value of $S$ around zero energy is enhanced to
$\pm 19.61 k_B/e$($\pm 1689.79 \mu V/K$).
Therefore, we can conclude that the characteristic features of
the conductance and thermopower at Fermi energy are strongly affected
by the disorder.


\begin{figure*}[tbh]
\par
\includegraphics[width=0.8\textwidth]{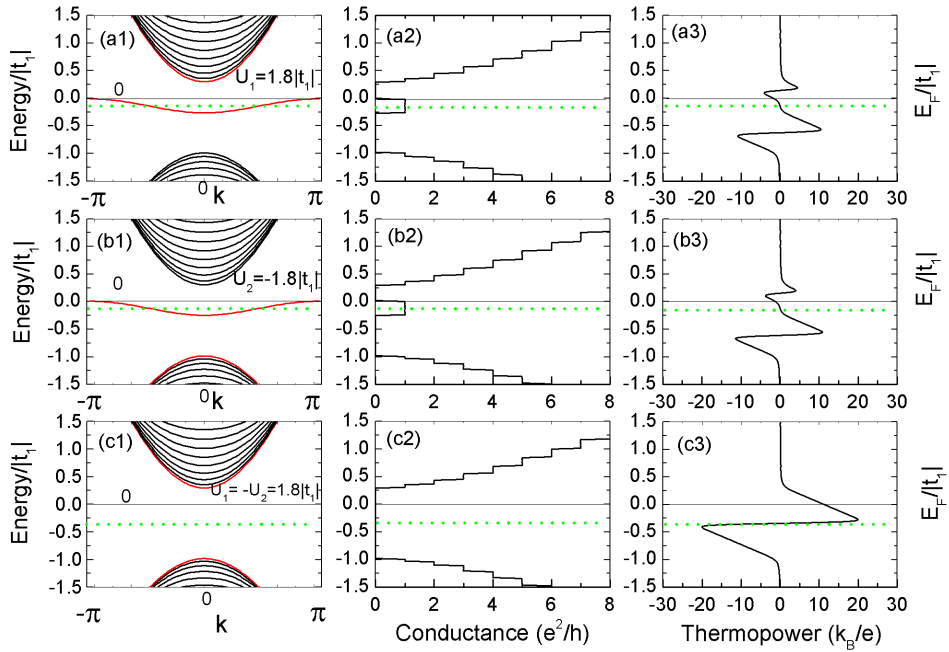}
\caption{(Color online) Calculated band structure, conductance and thermopower
of the ZPNRs in the presence of different boundary potentials at zero temperature.
(a) $U_{1}=1.8\vert t_1 \vert$, (b) $U_{2}=-1.8\vert t_1 \vert$, and
(c) $U_{1}=-U_{2}=1.8\vert t_1 \vert$.
Red curves represent the edge bands, and green lines
represent the Fermi energy. The width of the ZPNRs is set to $N=20$.
The temperature is taken to be $k_BT=0.03\vert t_1 \vert$ in the thermopower calculation.
} \label{fig.5}
\end{figure*}

In the following, we focus on how the band structure, conductance
and thermopower are affected by the boundary potentials for
the ZPNRs subject to a weak bias voltage.
The on-site energy is $U_{top}=\frac{1}{2}\Delta_g=0.01\vert t_1 \vert$ in the top layer
and $U_{bottom}=-\frac{1}{2}\Delta_g=-0.01\vert t_1 \vert$ in the bottom layer.
The boundary potentials $U_1$ and $U_2$ can also be adjusted
by potentials applied on the boundaries in both layers~\cite{Yao2009},
as shown in Fig.\ref{fig.1}(b).
When the boundary potentials $U_1$ and $U_2$ are adjusted in proper order,
the band structure, conductance and thermopower show significant properties.
In Fig.\ref{fig.5},
we first show the results of the band structure and conductance
at zero temperature.
As seen from Fig.\ref{fig.5}(a1), when only one boundary potential
 $U_{1}=1.8\vert t_1 \vert$ is applied to the top layer,
it is found that only one edge band
bends and fully merges into the bulk valence band.
Around zero energy, the first conductance plateau is strongly
reduced to $G=e^2/h$, as shown in Fig.\ref{fig.5}(a2).
Compared with that of the perfect ZPNRs, the metallic
nature decreases.
When only one boundary potential
$U_{2}=-1.8\vert t_1 \vert$ is applied to the bottom layer,
it is found
that only one edge band (the lower edge state)
bends and merges into the bulk conduction band, as shown in Fig.\ref{fig.5}(b1).
The behavior of the conductance is the same as in Fig.\ref{fig.5}(a2).

However, when the two boundary potentials $U_{1}$ and
$U_{2}$ are adjusted to $U_{1}=-
U_{2}=1.8\vert t_1 \vert$, both edge
bands bend and fully merge into the
bulk band, and the bulk energy gap is maximized,
as shown in Fig.\ref{fig.5}(c1). More interestingly, the quantized
conductance plateaus follow sequence $ G =n e^2/h $ with $n=0,2,3...$,
as shown in Fig.\ref{fig.5}(c2).
Around zero energy, a pronounced plateau with $G=0$ is found,
which can only be understood as due to the opening of the bulk
energy gap between the valence and conduction bands~\cite{Ma2011}.
Compared with that of the perfect ZPNRs, the conductance is strongly
suppressed, and the metallic characteristic disappears.

In the right side of Fig.\ref{fig.5}, we show the calculated thermopower
$S$ for different boundary potentials.
As we can see, when only one boundary potential of the top layer
$U_{1}$ is adjusted to $1.8\vert t_1 \vert$, the peak value of $S$
near zero energy is about $\pm 4.45 k_B/e$ ($\pm 383.45 \mu V/K$),
as shown in Fig.\ref{fig.5}(c1).
When the other boundary potential of the bottom layer $U_{2}$
is adjusted to $-1.8\vert t_1 \vert$, the behavior of $S$
is the same as the case for $U_{1}=1.8\vert t_1 \vert$.
In Fig.\ref{fig.5}(c3), when the two boundary potentials $U_{1}$ and
$U_{2}$ are adjusted to $U_{1}=-U_{2}=1.8\vert t_1 \vert$,
it is interesting to notice that $S$ has a much larger peak value
around zero energy. The peak value of $S$ is about
$\pm 19.95 k_B/e$($\pm 1719.09 \mu V/K$),
increasing more than twice, compared to the result of the perfect ZPNRs.
We attribute the large magnitude of $S$ to the opening of
the bulk energy gap near zero energy~\cite{Ma2011}.
Such an enhanced thermopower of the ZPNRs is
very beneficial for their thermoelectric applications.

\section{Summary}
\label{sec:sum}

In summary, we have numerically investigated the effect of edge
states on the conductance and thermopower in the ZPNRs based on the
tight-binding model and scattering-matrix method.
We find that the edge band structure, conductance, and thermopower
can be modulated by adjusting the bias voltage and boundary
potentials. Under the certain bias voltage $\Delta_g$,
the two-fold degenerate quasi-flat bands split completely,
one edge mode shifts upward and the other shifts
downward without changing their shapes. The first conductance
plateau around zero energy decreases to $0$, compared to that of
the perfect ZPNRs. Meanwhile, the thermopower around zero
energy increases.

On the other hand, when the boundary potentials are adjusted,
the band structure, conductance and thermopower show remarkable new
features. When only one boundary potential of the top layer $U_{1}$
or the bottom layer $U_{2}$ is applied, it is interesting to
find that only one edge band bends and merges into the bulk band.
Around zero energy, the first conductance plateau is strongly
reduced to $e^2/h$. However, when both boundary potentials
$U_{1}$ and $U_{2}$ are applied, both edge bands bend and
fully merge into the bulk band, and the bulk energy gap is maximized.
More interestingly, a pronounced plateau with $G=0$ is found
around zero energy, which can only be understood as due to the opening
of the bulk energy gap between the valence and conduction bands.
The conductance is strongly suppressed, and the metallic characteristic
disappears. Meanwhile, the thermopower is enhanced more than twice,
compared to the result of the perfect ZPNRs. We attribute the large
magnitude of $S$ to the opening of the bulk energy gap around zero energy.

\acknowledgments This work was supported by the National Natural Science Foundation
of China under grant numbers 11104146, 11574155, 11681240385 (R.M.),
11225420 (L.S.), 11174125, 91021003 (D.Y.X) and a project funded by the
PAPD of Jiangsu Higher Education Institutions. This work was also supported
by the State Key Program for Basic Researches of China under grant numbers
2015CB921202, 2014CB921103 (L.S.) and the Postdoctoral Science Foundation
of China under grant numbers 2014M551546, 2015T80532(R.M.).

\end{document}